\documentstyle{elsart}
\begin{document}
%\renewcommand{\baselinestretch}{1}

%My commands and substitutions:
\parskip 0pt
\parindent 15pt
\newcommand{\href}[2]{#2 ({\rm{#1}})}
\newcommand{\beq}{\begin{equation}}
\newcommand{\eeq}{\end{equation}}
\newcommand{\rl}{{r_{\rm L}}}
\newcommand{\calE}{{\varepsilon}}
\newcommand{\etal}{{et~al.}}
\newcommand{\const}{{\it const}}
\newcommand{\sfrac}[2]{{\textstyle\frac{#1}{#2}\,}}
\newcommand{\dg}{^{\rm o}}
%\lta and \gta produce < and > signs with twiddle underneath:
\def\spose#1{\hbox to 0pt{#1\hss}}
\def\lta{\mathrel{\spose{\lower 3pt\hbox{$\mathchar"218$}}
     \raise 2.0pt\hbox{$\mathchar"13C$}}}
\def\gta{\mathrel{\spose{\lower 3pt\hbox{$\mathchar"218$}}
     \raise 2.0pt\hbox{$\mathchar"13E$}}}

\begin{frontmatter}%---------------------------------------------------
\title{Region of magnetic dominance near a rotating black~hole}
\author[au,chalmers]{V.~Karas} and
\author[ktf]{M.~Dov\v{c}iak}
\address[au]{Astronomical Institute, Charles~University, \v{S}v\'edsk\'a~8,
 \hbox{CZ-150\,00~Prague,} Czech~Republic}
\address[chalmers]{Department of Astronomy and Astrophysics, G\"oteborg
 University and Chalmers University of Technology,
 \hbox{S-412\,96~G\"oteborg,} Sweden}
\address[ktf]{Department of Theoretical Physics, Charles~University,
 V~Hole\v{s}ovi\v{c}k\'{a}ch~2, CZ-180\,00~Prague, Czech~Republic}

\vspace{4cm}
\begin{flushleft}
{\normalsize \underline{Running head:} Magnetic dominance near black holes}

\vspace{2cm}

{\normalsize
\underline{Postal address for correspondence:}

V. Karas\\
Astronomical Institute\\
\v{S}v\'edsk\'a~8\\
CZ-150\,00 Prague\\
Czech~Republic

\underline{Electronic mail:} karas@mbox.cesnet.cz

\underline{Fax:} +420-2-21911292

\underline{Telephone:} +420-2-540395, +420-2-535764
}
\end{flushleft}

\newpage

\begin{abstract}%------------------------------------------------------
\noindent
Processes of collimation of electrically charged particles near a
rotating black hole are discussed. It is assumed that the black hole is
immersed in a weak magnetic field aligned with rotation axis. This
situation is relevant for understanding pre-collimation of astrophysical
jets. Magnetic field affects motion of material and restricts
validity of various scenarios which adopt the test-particle (cold
plasma) approximation. A simplified criterion which estimates
relevance of such approximation is discussed in connection with
the mechanism of the dissipative collimation, as proposed by de~Felice \&
Curir \cite{fc92}.
\end{abstract}%--------------------------------------------------------

%---------------------------------------------------------

\end{frontmatter}%-----------------------------------------------------

\section{INTRODUCTION}%------------------------------------------------
\noindent
Numerous observations confirm that collimated outflows of matter are
rather generic phenomenon connected with certain types of astronomical
objects \cite{h91}. These jets exist on different length-scales and they
are associated with various types of sources ranging from stars to
galactic nuclei---i.e.\ over nine orders of magnitude in the mass of the
central source. The origin of jets is probably diverse but they share
common properties. For example, it has been speculated about analogies
between electromagnetic processes which accelerate particles near
pulsars \cite{b84,m82} versus processes in magnetospheres of
supermassive black holes in active galactic nuclei (AGN)
\cite{bz77,ntt91}. AGNs with jets show many diverse properties, but it
is tempting to link the differences, at least partially, to orientation
of these objects with respect to the observer. The interest in this
subject has been amplified by recent discoveries of relativistic
outflows in our Galaxy \cite{hr95,mr94,tjp95} which have their
well-known counterpart in extragalactic superluminal jets \cite{zp90}.
Unifying schemes have been proposed for cosmic sources with jets. Useful
review articles summarizing our knowledge can be found in the
literature, both for stellar-scale objects in the Galaxy \cite{plg91}
and for extragalactic jets \cite{bbr84,up95}.

Extragalactic jets are presumably formed in the innermost regions of the
source (within a few or a few tens of gravitational radii, $R_g$, from
the center) and they emanate outwards along rotation axis of
the central object. We assume the model with a compact rotating object
in the center. Gravitational field of the central object is approximated
by the Kerr metric while self-gravity of the jet material and
accretion disk is neglected. It has been widely recognized that
large-scale outflows can be adequately described in the
magnetohydrodynamic regime but situation very close to the horizon is
less understood. The initial phase of the jet formation is sometimes
called {\em pre-collimation}, in distinction to the processes of
successive collimation which operate in more distant regions.

Several mechanisms of the jet pre-collimation have been proposed.
Magnetic fields play most probably a major role in focusing and
maintaining collimated outflows on their course. Numerous authors have
studied  collimation within the hydromagnetic framework
\cite{bp82,c86,cb92,lmm86,ops97}. It has been shown that toroidal
component of the magnetic field is maintained by rotation of the
accretion disk and enhances collimation \cite{ukr95}. We were interested
in the contribution of the Kerr geometry to the resulting collimation of
jets, and it appears quite natural to presume that our attention can be
restricted to the region within a few tens of gravitational radii from
the center. The inner region of the object is crucial for the theory of
jet formation but it remains beyond current observational capabilities
which are of the order of $(10^2$--$10^3)\,R_g$ for extragalactic
sources (the best linear resolution, approximately 0.01~pc, has been
achieved with the radio jet in the galaxy 3C\,274; \cite{jb95}).

Abramowicz \& Piran \cite{ap80} and Sikora \& Wilson \cite{sw81}
consider collimation inside a funnel of a luminous thick accretion disk.
Material of the jet is in mutual interaction with the disk radiation
which determines its terminal speed \cite{p87}. Nowadays, the idea of
extremely thick disks with very narrow funnels and highly
super-Eddington luminosity (e.g., \cite{fkr95}) is not favoured because
these models suffer from several inconsistencies but the general scheme
of jets flowing along the disk axis remains viable with more
sophisticated models of accretion disks. The model has been advanced by
detailed quantitative calculations of acceleration/deceleration of jets
due to radiation pressure and losses due to cooling, both within the
framework of the hydrodynamic \cite{fb95,lcb92,pe96} and the test-particle
approximation \cite{mk89,ssb96,sw81,vk91}. These calculations impose
strong limitations on radiatively driven jets because their material
cannot reach Lorentz factors much greater than unity. Hydromagnetic
scenario is thus currently favoured \cite{b95,ops97}.

The above-mentioned schemes consider a massive compact object to be
present in the center of the source but the Kerr geometry and the
effects of general relativity are not crucial for their function. On the
other hand, de~Felice \& Curir \cite{fc92} and de~Felice \& Carlotto
\cite{fc95} have explored a special family of geodesics spiralling along
the axis of the Kerr black hole (vortical geodesics; see
\cite{fc72,fc90}) and determined constraints on the rate of change of
energy and angular momentum of outflowing material that may result in
collimation. They did find collimation but the physical nature of
dissipative processes that cause the loss of particle energy and angular
momentum remains unclear. It is the aim of the present contribution to
advance the latter model by systematic discussion of the parameter space
of particle trajectories. We want to determine, by a simplified but
systematic approach, circumstances under which the above-mentioned
scheme could be relevant (rather than build up our own model). In other
words, we ask whether specific features of motion in Kerr geometry are
relevant for pre-collimation of astrophysical jets. Indeed, the original
motivation to deal with this problem was a suspicion that even
relatively weak magnetic fields impose strong limitations upon the
model. (This fact has been quoted and applied in numerous works; we wish
to discuss the problem in a more systematic way with the Kerr geometry.)
Our arguments could be applied also to other models based on the
properties of the geodesic motion \cite{bsh93}, spin-curvature coupling
(another purely general relativistic effect, \cite{s97}), and influence
of magnetic fields which affect the spacetime geometry \cite{kv90}. The
latter process requires dimensionless product $\beta$ of magnetic
intensity and the mass of the central object to be of the order of
unity, which is value much higher than
$\beta\doteq5\times10^{-8}B_4M_8\approx10^{-8},$ where $B_4\equiv
B/(10^4\,$gauss) and $M_8\equiv M/(10^8\,M_\odot)$ denote typical values
relevant for the innermost magnetosphere in AGNs. Direct interaction of
the jet material with external electromagnetic fields appears more
important. (Geometric units will be used hereafter, $c=G=1;$ time, mass,
electric charge and [magnetic intensity]$^{-1}$ have dimension of
length, i.e.\ cm. Next, quantities with dimension of length will be
expressed in units of $M$.)

Structure of this text is as follows: First, equations for generalized
energy and generalized angular momentum of electrically charged
particles are given. Dimensionless parameters characterizing, locally,
electromagnetic effects upon the test-particle motion are then defined
and evaluated. It should be noted that in the present contribution we do
not address the problem of extraction of rotational energy from a
magnetized black hole which also relies on properties of the Kerr
spacetime \cite{htn92,o92,tnt90,wd89}, neither we discuss purely
electromagnetic collimation due to toroidal magnetic fields.

\section{CHARACTERISTIC LENGTH-SCALE}%---------------------------------
\subsection{Details of the model}%.....................................
\noindent
We assumed that a weak magnetic field is generated far from the black
hole, outside the region where motion of matter is studied. The
spacetime is described by the Kerr metric with an asymptotically uniform
magnetic test field \cite{w74}. This solution reflects the large-scale
field which is generated far from the black hole. We further assumed
that the field is aligned with rotation axis because non-aligned fields
exert torque on the black hole, so that its rotation slows down and
the black hole gets aligned with the field \cite{kl77,pb77,p72}. Typical
time-scale for the black-hole alignment due to electromagnetic torques
is proportional to $\beta^{-2}$ and rather long (in physical units,
$\tau/(10^{10}\,\mbox{yr})\approx M_8^{-1}B_4^{-2}$) but tidal
effects enhance alignment and decrease $\tau$ \cite{kp85,sf96}.

Structure of an asymptotically uniform magnetic field is simple and
collimation by Kerr geometry is presumably easier to notice \cite{s95}.
(As mentioned above, toroidal fields, $B_\phi$, contribute to purely
electromagnetic collimation and to the total power-output from a black
hole \cite{pv94}; in time-dependent calculations, $B_\phi=0$
is often taken as an initial configuration of the magnetic field
\cite{ops97}). Jets are formed near a black hole but the
material which forms jets is an open question (see, e.g., \cite{b95,k85}
for general issues related to formation of jets). In the case of
electron-positron plasma, the specific electric charge of individual
particles is $|\tilde{q}|\approx2\times10^{21}$ (elementary charge $q$
and mass of electron $m$ will be assumed in numerical estimates,
$\tilde{q}\equiv q/m$). Dimensionless parameter
$\epsilon\equiv|\beta\tilde{q}|$ determines motion of charged particles.

Trajectory of individual particles is determined by Larmor gyrations
around magnetic lines of force (Larmor radius $\rl$) and the drift
motion across the field-lines (e.g., \cite{af63}). Motion of
electrically charged particles near a black hole has been studied by
numerous authors \cite{bsb89,d86,gp78,s79}. Prasanna \cite{p80} and
Karas \& Vokrouhlick\'y \cite{kv90} show examples of test-particle
trajectories near magnetized black holes. Now we are interested in the
bulk motion of the material rather than individual trajectories which
can be very complicated. It is thus useful to disregard local gyrations
and accept the guiding-center approximation. Damour \etal\ \cite{dhr78}
determined shape of the plasma flow-lines near a weakly magnetized Kerr
black hole in the guiding-center approximation. Discussion has been
generalized to the case of an electrically charged rotating black hole
by Hanni \& Valdarnini \cite{hv79} (with a weak magnetic field), and
Karas \& Vokrouhlick\'y \cite{kv91} (an exact solution of the
Einstein-Maxwell equations with an arbitrarily strong magnetic field).
These authors verified that plasma moves along rotation axis and they
concluded that asymptotically uniform magnetic field evidently
contributes to collimation.

The situation becomes much less understood when mechanism proposed by
de~Felice \& Curir \cite{fc92} (motion along vortical geodesics plus
tiny, yet undetermined dissipation of energy and angular momentum) is
taken into account. One can, however, estimate relevance of this scheme
by calculating the relative change of orbital parameters due to magnetic
field.

\subsection{Rate of change of orbital parameters}%.....................
\noindent
In this section, we will estimate the rate of change of orbital
parameters (energy and angular momentum with respect to rotation axis)
of individual particles and integrate the rate over particle distribution.
This approach gives us a {\em local criterion\/} of importance of
electromagnetic forces.

The model is described by an axisymmetric stationary spacetime metric
$g_{\mu\nu}$ $(\mu,\nu=0,\ldots,3).$ Electromagnetic test field is
characterized by the tensor of electromagnetic field, and by
corresponding four-potential: $F_{\mu\nu}\equiv A_{\nu,\mu}-A_{\mu,\nu}.$
It should be emphasized that we do not have to impose further restrictions
upon gravitational and electromagnetic fields, apart from the above mentioned
assumption about axial symmetry and stationarity. To be specific,
however, we will consider the Kerr metric \cite{mtw73} with an aligned
asymptotically uniform magnetic field. Structure of the magnetic field
has been explored by several authors (e.g., \cite{hr76,pb77} in the case
of aligned fields, and \cite{bj85,k77,k89} in the case of inclined
fields). In Boyer-Lindquist coordinates,
$x^\mu\equiv\{t,r,\theta,\phi\}$,
\beq
A_t=\beta a\left[r\Sigma^{-1}\left(1+\cos^2\theta\right)-1\right],
\label{at}
\eeq
\beq
A_\phi=\beta \sin^2\theta\left[\sfrac{1}{2}\left(r^2+a^2\right)-
 a^2r\Sigma^{-1}\left(1+\cos^2\theta\right)\right].
\label{ap}
\eeq
Here, $\Sigma\equiv r^2+a^2\cos^2\theta.$  Each timelike
geodesic (four-momentum $p^\mu\equiv mu^\mu$) is associated with two
conserved quantities---specific energy
\beq
\calE_0\equiv-u_{t\,\mid\,q=0},
\label{en0}
\eeq
and specific angular momentum with respect to the symmetry axis
\beq
\lambda_0\equiv u_{\phi\,\mid\,q=0}.
\label{el0}
\eeq
Four-momentum along a trajectory
of an electrically charged particle is determined by equation
\beq
\frac{{\rm D}u^\mu}{\rm D\tau}=\tilde{q}\,F^\mu_\nu u^\nu,
\eeq
and corresponding conserved quantities are:
\beq
\calE\equiv-\left(u_t+\tilde{q}A_t\right),\quad
\lambda\equiv u_\phi+\tilde{q}A_\phi
\eeq
(generalized energy and angular momentum component, respectively).

Any scenario which explains collimation of particles in terms of
curvature effects acting upon free particles (as in Bi\v{c}\'ak,
Semer\'ak \& Hadrava \cite{bsh93}) or particles that are slightly
perturbed by dissipative forces (as in de~Felice \etal\
\cite{fc95,fc92}) turns out to be astrophysically irrelevant when
electromagnetic forces strongly affect motion of particles. In order to
introduce a quantitative criterion for electromagnetic effects, one can
define two parameters, $\delta\calE(r,\theta;p^\mu)$ and
$\delta\lambda(r,\theta;p^\mu)$, which characterize the relative change
of $\calE_0$ and $\lambda_0$:
\beq
1-\frac{\calE_{0\,\mid\, x^\mu+u^\mu\d\tau}}{\calE_{0\,\mid\, x^\mu}} =
 \tilde{q}\,\Lambda\,\frac{F_{tr}u^r+
 F_{t\theta}u^\theta}{\tilde{q}A_t+\calE}\d\sigma
 \equiv\delta\calE\,\d\sigma,
\label{de}
\eeq
\beq
1-\frac{\lambda_{0\,\mid\, x^\mu+u^\mu\d\tau}}{\lambda_{0\,\mid\, x^\mu}} =
 \tilde{q}\,\Lambda\,\frac{F_{r\phi}u^r+
 F_{\theta\phi}u^\theta}{\lambda-\tilde{q}A_\phi}\d\sigma
 \equiv\delta\lambda\,\d\sigma.
\label{dl}
\eeq
Here, $\d\sigma$ denotes interval of proper time scaled with the
light-crossing time across the characteristic length-scale $\Lambda$:
$\d\sigma\equiv\d\tau/\Lambda.$

It is further postulated that $\Lambda\propto\rl=\Gamma
v\tilde{q}^{-1}\beta^{-1}$ (with $\Gamma=1/\sqrt{1-v^2}\lta10$ denoting
the Lorentz factor, $v$ velocity with respect to the locally
non-rotating frame). Hence, $\langle\delta\calE\rangle$ and
$\langle\delta\lambda\rangle$ become independent of magnetic field
strength. This fact is understandable: the Larmor radius decreases with
$\beta$ increasing, which means that the characteristic length along
which the change of parameters is determined in
eqs.~(\ref{de})--(\ref{dl}) decreases as well. Such a choice of
$\Lambda$ is well-founded (for our purpose of an order-of-magnitude
criterion) because geodesic trajectory certainly cannot approximate a
real trajectory of a charged particle on a scale greater then $\rl.$ (On
the other hand, it is easy to evaluate $\langle\delta\calE\rangle$,
$\langle\delta\lambda\rangle$ also for a different choice of $\Lambda$;
$\beta$ then becomes another free parameter, however.) Considering the
collimation processes acting on length-scales of a few $R_g$ (as in
de~Felice \& Curir \cite{fc92}), $\Lambda$ should be comparable with or
greater than $R_g$; this is further restriction on the upper limit for
$\beta.$

Both parameters, $\delta\calE$ and $\delta\lambda$, are defined locally
(i.e.\ $r,$ $\theta$ given) and they also depend on particle's $p^\mu.$
We study the region outside the black-hole horizon by averaging over
distribution of particles in the momentum space. We define

\beq
\langle\delta\calE\rangle^2\equiv\frac{1}{4\pi}\int_{4\pi}\d\Omega
\int_{\Gamma_{\rm min}}^{\Gamma_{\rm
max}}\d\Gamma\,n(\Gamma)\,\delta\calE^2, \label{ave}
\eeq
and analogously
\beq
\langle\delta\lambda\rangle^2\equiv\frac{1}{4\pi}\int_{4\pi}\d\Omega
\int_{\Gamma_{\rm min}}^{\Gamma_{\rm
max}}\d\Gamma\,n(\Gamma)\,\delta\lambda^2. \label{avl}
\eeq

Integration is taken over the particle distribution in energy,
$n(\Gamma),$ and over all directions of their local velocity.
Values of $\langle\delta\calE\rangle\gta1,$
$\langle\delta\lambda\rangle\gta1$ mean that approximation of geodesic
motion with a small perturbation is inappropriate, while values of both
parameters much less than unity, {\it simultaneously\/} with
$\Lambda\gta R_g$, indicate that this approximation might be meaningful.

\begin{figure}
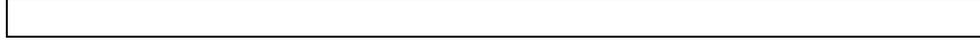

%\epsfxsize=\hsize
%\epsfbox{fig1r.ps}
\framebox{\hspace*{0.9\hsize}}
\caption{Graphs of $\langle\delta\calE\rangle$ [panels (a), (c)] and
$\langle\delta\lambda\rangle$ [panels (b), (d)] as a function of radius
$x\equiv1-R_g/r$ $(R_g=1+\protect\sqrt{1-a^2}, 0<x<1)$ and $\theta$
$(0<\theta<\pi).$ Upper panels,
(a)--(b), show mean values taken over all trajectories with
$\Gamma\leq\Gamma_{\rm max}$ while lower panels, (c)--(d), deal with
vortical trajectories only (notice a narrow gap near $\theta=\pi/2$:
vortical trajectories do not cross equatorial plane). Here, $a=1,$
$0\lta\Gamma\lta3,$ $s=0.$ See the text for details.
\label{fig1}}
\end{figure}

\begin{figure}
%\epsfxsize=\hsize
%\epsfbox{fig2r.ps}
\framebox{\hspace*{0.9\hsize}}
\caption{As in Fig.~\protect\ref{fig1} but for $0\lta\Gamma\lta2,$ $s=2.$
\label{fig2}}
\end{figure}

\section{RESULTS}%...................................................
\noindent
We evaluated parameters $\langle\delta\calE\rangle$ and
$\langle\delta\lambda\rangle$ for the Kerr metric and electromagnetic
test field (\ref{at})--(\ref{ap}). Equations (\ref{en0})--(\ref{el0})
can be written in the explicit form:
\beq
\calE_0=\sqrt{\Sigma\Delta A^{-1}}\,\Gamma+\omega\lambda_0,
\eeq
\beq
\lambda_0=\sqrt{A\Sigma^{-1}}\,\Gamma v^\phi\sin^2\theta,
\eeq
where $\Delta=r^2-2r+a^2,$ $A=(r^2+a^2)^2-\Delta a^2\sin^2\theta,$
$\omega=2arA^{-1}$ are functions from the Kerr metric in standard
notation \cite{mtw73}; $v^\phi$ is azimuthal component of the speed of
particle with respect to LNRF. Energy distribution of particles was
approximated by a power-law: $n(\Gamma)\propto\Gamma^{-s}$ in a
restricted interval of energy ($\Gamma_{\rm max}\lta20,$ $0\lta
s\lta2;$ power-law energy distribution is motivated by astrophysically
relevant situations).

Figures \ref{fig1}--\ref{fig2} illustrate our results for an isotropic
distribution of particles with respect to LNRF. We assumed
$\Lambda\approx\rl$ (characteristic length) for definiteness. In these
two figures, $\langle\delta\calE\rangle\ll1,$
$\langle\delta\lambda\rangle\ll1$ [panels (a) and (b)], and one
concludes that approximation of the close-to-geodesic motion may be
relevant for modelling the jet precollimation in the given region of
$(r,\theta)$. We have also examined separately the case of vortical
trajectories which play a crucial role in discussion of de~Felice \&
Curir \cite{fc92} [panels (c) and (d)]. In the latter case, the
assumption about isotropic distribution is supplemented by specific
conditions for vortical trajectories:

\beq
\tilde{\Gamma}>0,\quad |L|\leq a^2\tilde{\Gamma},\quad
 L<\lambda_0^2\leq\frac{L+a^2\tilde{\Gamma}}{4a^2\tilde{\Gamma}}
\eeq
($L$ denotes the fourth constant of the Kerr metric,
$\tilde{\Gamma}\equiv\calE_0^2-1$). Regions in the $(r,\theta)$ plane
with $\langle\delta\calE\rangle\gta1,$
$\langle\delta\lambda\rangle\gta1$
have been excluded from Figures for clarity (particularly, close
to the horizon and rotation axis).

Figures \ref{fig1}--\ref{fig2} represent a typical situation.
Further illustrations in which the parameter space is investigated
systematically can be found on World-Wide Web.\footnote{
http://otokar.troja.mff.cuni.cz/user/karas/\-au\_www/\-karas/\-papers.\-htm
}

\section{CONCLUSION}%...................................................
\noindent
We derived an order-of-magnitude criterion for possible relevance of
those models of the jet pre-collimation (at distances of few $R_g$ from
the black hole) which are based upon test-particle approximation:
$\langle\delta\calE\rangle^2$, $\langle\delta\lambda\rangle^2\ll1$
(independently of $B$), and $\Lambda\gta R_g$ (depends on the value of
$B$: $\Lambda\propto B^{-1}$). Our estimates employ characteristic
length-scales which can be shorter when collisions of particles in
plasma are important or if the magnetic field is dominated by a
short-scale chaotic component. This means the region of magnetic
dominance can be larger than our criterion indicates. Although only
rough estimates of strength of the magnetic field are currently
available, it appears that the guiding-center approximation is the most
relevant approach among the models based on test-particle motion. One
should note that general relativity effects remain important for motion
of material not only because of the presence of the black hole in the
center but also because the structure of electromagnetic field itself is
affected by strong gravity. It is also worth noticing that explicit
formulae for non-aligned test fields are known and can be studied in a
similar way.

Our present discussion has been restricted by assumptions about the
large-scale structure of the magnetic field, isotropic distribution of
particle velocities in LNRF, and power-law distribution of their energy.
We do not expect our results to be sensitive to these assumptions unless
the particle distribution is very anisotropic; it should be repeated
that farther from the center hydromagnetic description is appropriate.

\bigskip

\noindent
{\bf ACKNOWLEDGEMENT}
\smallskip\par\noindent
The authors acknowledge helpful comments and critical remarks from
professor Fernando de Felice, and from participants of the Seminar on
General Relativity in Prague. We acknowledge suggestions by an anonymous
referee concerning definition of $\Lambda$. We thank Joan Evans for
reading the manuscript. This work has been supported by the grants GACR
205/\linebreak[2]97/\linebreak[2]1165 and GACR
202/\linebreak[2]96/\linebreak[2]0206 in the Czech Republic.

%{\bf References}{}%----------------------------------------------------

\end{document}